\documentclass[a4paper,11pt]{article}

\pdfoutput=1 
\usepackage{cite}
\usepackage{graphicx}
\newcommand*{\email}[1]{\href{mailto:#1}{\nolinkurl{#1}} } 
\usepackage{hyperref}
\usepackage[affil-it]{authblk}
\usepackage{enumerate}

\usepackage{newtxtext}

\usepackage{slashed}
\usepackage{wrapfig}
\usepackage{textcomp}
\usepackage{amsmath,amssymb,fontenc, times}
\usepackage{float}
\usepackage{verbatim}

\usepackage{bm}
\usepackage{esvect}

\usepackage{diagbox}
\usepackage{adjustbox}
\usepackage{multirow}

\newcommand{\be}{\begin{equation}}
\newcommand{\ee}{\end{equation}}
\newcommand{\bea}{\begin{eqnarray}}
\newcommand{\eea}{\end{eqnarray}}
\newcommand{\bes}{\begin{subequations}}
\newcommand{\ees}{\end{subequations}}
\newcommand{\bear}{\begin{equation}\begin{array}}
\newcommand{\eear}[1]{\end{array}\label{#1}\end{equation}}

\newcommand{\beg}{\begin{equation}\begin{gathered}}
\newcommand{\eeg}{\end{gathered}\end{equation}}
\newcommand{\beal}{\begin{equation}\begin{aligned}}
\newcommand{\eeal}{\end{aligned}\end{equation}}
\newcommand{\begg}{\begin{gather*}}
\newcommand{\eegg}{\end{gather*}}

{\end{list}}
\newcounter{enumct}

\usepackage{color}    
\usepackage{lipsum}
\usepackage{soul}
\definecolor{darkgreen}{RGB}{11,150,35}

\usepackage{ulem}

\begin {document}

\title{An effective gauge field theory of the  nucleon interactions}

\author[a,b]{Eduard Boos \footnote{\email{boos@theory.sinp.msu.ru, Eduard.Boos@cern.ch}}}

\affil[a]{Skobeltsyn Institute of Nuclear Physics, Lomonosov Moscow State University, Leninskie Gory, 119991, Moscow, Russia}
\affil[b]{Faculty of Physics, Lomonosov Moscow State University, Leninskie Gory, 119991, Moscow, Russia}

\maketitle


\begin{abstract}
We discuss the possibility of constructing an effective gauge field theory of the nucleon interations based on the ideas of isotopic invariance  as well as hypercharge invariance as a local gauge symmetry and spontaneous breaking of this symmetry. The constructed effective field theory predicts the structure of interactions of protons and neutrons with $\rho$- and $\sigma$-mesons, with pi-mesons and photons, as well as interactions of these particles with each other. The Lagrangian of the theory consists of several parts parts involving dimension 4 and 5 gauge invariant operators. Feynman rules for physical degrees of freedom as follow from the Lagrangian  define the structure of diagrams for one-boson exchanges between nucleons predicting the internucleon one-boson exchange potential as well as nucleon scattering amplitudes. The range of applicability of the effective theory is discussed and estimates are made of the resulting coupling constants. The theory predicts the mass of the neutral $\rho^0$-meson to be about $1\,MeV$ larger than the mass of the charged mesons $\rho^{\pm}$. The vector $\omega$-meson, which is a sterile particle with respect to the considered gauge group $SU_I(2)\times U_Y(1)$, can be added to the scheme by means of a gauge-invariant operator of dimension 5, as shown in Appendix ~A.
\end{abstract} 

\newpage
\tableofcontents
\section{Introduction}
One of the first theories of nucleon-nucleon interactions was the Yukawa theory \cite{Yukawa:1935xg}.
Subsequently, the meson exchange potentials between nucleons began to be intensively constructed and studied. Various variants of potentials were obtained, which made it possible to understand many properties of nuclei, to quantitatively describe data on cross sections, phases, and scattering lengths. 
A number of specific potentials have received much attention, such as the Moscow potential \cite{Neudatchin:1975zz}, the Argonne potential \cite{Wiringa:1994wb} or later ones built on the basis of field theories, such as the Bonn potential \cite{Machleidt:2000ge}, the Nijmegen potential \cite{Schulze:2011zza, Schulze:2013jka} and others. Review article \cite{Machleidt:2022kqp} provides a detailed overview of the history and current state of research on internucleon potentials, in particular,  the internucleon one-boson-exchange potentials (OBEP).

The idea of exchange interaction, the exchange of pi-mesons and other mesons with masses less than the mass of the nucleon, turned out to be exceptionally fruitful not only for describing nuclear forces. Yukawa-type interactions formed the basis for describing the interaction of leptons and quarks with the Higgs field in the Standard Model (SM)\cite{Glashow:1961tr, Weinberg:1967tq, Salam:1968rm}. \footnote{When constructing the effective field  theory, we will use the notation similar to the notation adopted in 
\cite{Boos:2014vpa, Boos:2015lbf} when presenting the electroweak part of the SM.} This made it possible to assign masses to fermion fields, predict the strength of fermion interactions with the Higgs boson, and introduce the Cabibbo-Kobayashi-Maskawa mixing matrix. Non-Abelian gauge bosons and interactions due to the exchange of these bosons were introduced by Yang and Mills \cite{Yang:1954ek} when considering Heisenberg's isotopic symmetry as a local symmetry. An attempt to interpret the emerging gauge fields, called Yang-Mills fields, as fields of vector $\rho$-mesons could not be successful due to the massless nature of the gauge fields. In the present  paper, we continue research in this direction, using an analogue of the Braut-Engler-Higgs (BEH) \cite{Englert:1964et, Higgs:1964ia, Higgs:1964pj} mechanism of spontaneous symmetry breaking in the SM  to impart mass to gauge fields.

The part of the SM that describes strong interactions, quantum chromodynamics (QCD), is a theory of the interaction of quarks and gluons, which follows from the requirement of gauge invariance with respect to the $SU(3)_c$ group. Many results of experiments in various ranges of characteristic distances or energies are successfully explained in terms of perturbative or nonperturbative QCD. From the point of view of QCD, the interactions between hadrons, as colorless composite objects of quarks and gluons, are some kind of effective interactions, reminiscent in a sense of the van der Waals forces between extended neutral objects in electrodynamics (QED). Of course, the characteristic behavior of such internucleon interactions or forces acting between nucleons differs significantly from the van der Waals interactions in QED. The carriers of fundamental strong interactions, gluons, carry color and, due to confinement, cannot propagate outside colorless hadrons. And the interaction between hadrons can be carried out through the exchange of only colorless objects, which are mesons.

It is very difficult to construct a complete theory of internucleon interactions following from the fundamental QCD Lagrangian. The perturbation theory with respect to the QCD coupling constant is obviously inapplicable in this case, and nonperturbative methods are not fully developed for such a problem. However, in the case when the characteristic energies are less than the characteristic inverse size of hadron $\frac{1}{10^{-13}\, cm} \approx \,\,200\,MeV$, one can construct effective field theories describing internucleon interactions based on the use of symmetries motivated by the symmetries of the fundamental Lagrangian.

One of such symmetries, which follows from QCD neglecting the current masses of quarks, is the global chiral symmetry corresponding to the $SU(2)_L\times SU(2)_R$ transformation group of left- and right-chiral quark doublets. In the Standard Model, quarks acquire masses through the BEH mechanism, quark condensate appears in complex QCD vacuum. As a result the $SU(2)_L\times SU(2)_R$ symmetry is broken, and the resulting pseudoscalar Goldstone bosons are pi-mesons with small masses of the order of 100\,MeV compared to the masses of protons and neutrons. The theory of pion-nucleon and pion-meson forces was developed by Steven Weinberg \cite{ Weinberg:1968de}. In this approach the vector mesons appear as dynamical fields
realising a gauge symmetry non-linearly with a number of attractive phenomenological features \cite{Coleman:1969sm, Callan:1969sn, Bando:1984ej, Meissner:1987ge, Bando:1987br, Harada:1992np, Tanabashi:1993np, Birse:1996hd, Gamermann:2006nm, Wu:2010jy} (see also review 
\cite{Machleidt:2011zz} and references therein and to it). Masses of the gauge fields 
are introduced by means of non-abeliean variant of the Stueckelberg mechanism 
\cite{Stueckelberg:1938hvi, Kunimasa:1967zza, Slavnov:1972qb}.

The symmetry $SU(2)_L\times SU(2)_R$ is broken to the vector symmetry $SU(2)_V$. This symmetry is nothing more than isotopic symmetry, introduced at one time Heisenberg\cite{Heisenberg:1932dw}.  On the basis of this symmetry, the ideas of the non-Abelian gauge invariance of Yang and Mills, and the idea of spontaneous symmetry breaking, we will construct an effective field theory of nucleon and meson interactions. The linear realization of spontaneous symmetry breaking is used similarly to the electroweak Standard Model. 
The constructed gauge invariant effective theory predicts a very specific structure of the interaction vertices of not only protons and neutrons with rho mesons, sigma mesons, pions and photons but also the self-interaction of vector gauge bosons, their interaction with scalar sigma and pseudoscale pi mesons. The Feynman rules, following from the Laragangian of the effective theory, allow to calculate from the presented Feynman diargrams single-boson exchange potentials, amplitudes, scattering cross sections, etc. using well-known procedures. These calculations and corresponding
phenomenological analyses are out of the scope of the current theoretical study and are planned as the next step.

An interesting prediction of the theory is the small difference in masses between charged and neutral $\rho$ mesons, similar to the difference in masses of $W^{\pm}$ and $Z$ bosons in the Standard Model. This difference turns out to be on the order of current experimental errors, which motivates future more accurate measurements of the rho meson masses.

The paper is organized as follows. In Section 2 we construct a gauge-invariant Lagrangian based on the hypothesis that isotopic invariance is a gauge symmetry. The masses of the gauge bosons are obtained as a result of the spontaneous breaking of this symmetry due to the interaction with the introduced complex scalar doublet, which has a nontrivial vacuum expectation value. A scalar boson arises, similar to the Higgs boson in the SM, and acquires mass due to the interaction with the nontrivial condensate. This boson is interpreted as a $\sigma$-meson. Section 3 shows how one can introduce a gauge-invariant Lagrangian of dimension 5, which consists of two operators and ensures the interaction of the $\sigma$-meson with the fermionic fields of protons and neutrons. In this case, the gauge invariance allows the appearance of different corrections to the masses of protons and neutrons. In Section 4 the Lagrangian in elaborated  under the assumption that the gauge symmetry includes not only the isotopic symmetry, but also the hypercharge symmetry. By analogy with the electroweak part of the SM, as a result of the spontaneous symmetry breaking, only one unbroken symmetry remains - the electromagnetic one, and the resulting Lagrangians of charged and neutral nucleon currents are discussed. The massive vector bosons are interpreted as the $\rho$-mesons. Section 5 is devoted to the gauge-invariant inclusion of interactions with pi-mesons as triplets of the isotopic spin group, but having no hypercharge. In Section 6 the determination of a part of the parameters of the obtained Lagrangians is discussed, in particular, estimates of the vacuum expectation value of order 260 $MeV$ and the value of the gauge interaction constant. In conclusion  the main obtained results are given including the possibility of gauge-invariant inclusion in the scheme of the vector $\omega$-meson, which is a sterile particle with respect to the gauge group of isotopic spin and hypercharge. Corresponding interaction Lagrangian is given in Appendix A.

\section{Isotopic invariance as a gauge symmetry}

Following the original idea of Yang and Mills \cite{Yang:1954ek}, we assume that the isotopic symmetry
$SU(2)_I$ is a local gauge symmetry. We also assume that the field of the nucleon
  $\Psi_N(x) = \left(
\begin{array}{c}
\psi_p \\
\psi_n
\end{array}
\right)$
is an isotopic doublet and is a local spinor field that satisfies the Dirac equation
\begin{equation}
\label{Dirac-eq}
  (\mathbf{i}\partial_{\mu}\gamma^{\mu} - M_N)\Psi_N = 0,
\end{equation}
where $\gamma^{\mu}$ are the Dirac gamma matrices, the upper $\psi_p$ and lower $\psi_n$ components of the doublet are the fields of the proton and neutron with spin 1/2, isospin 1/2, and 
isospin projections $T^3_{\psi_{p,n}} = \pm 1/2$, $M_N$ is a  bare nucleon mass.

In order that the Lagrangian leading to the Dirac equation (\ref{Dirac-eq}),
    \begin{equation}
\label{Fermion-lagr}
  L_{\Psi} = \bar\Psi_N (\mathbf{i}\partial_{\mu}\gamma^{\mu} - M_N^0)\Psi_N
\end{equation}
did not change under local rotation in the isotopic space
   \begin{equation}
\label{gauge-tranf_spinor}
  \Psi_N(x) \rightarrow \Psi^\prime_N(x) = U(x)\Psi_N(x), \, \, U(x)= e^{\mathbf{i} \alpha_j(x)\tau_j}
\end{equation}
where $\tau_j=\sigma_j/2$ are the generators of the $SU_I(2)$ group, $\sigma_j$ are the Pauli matrices,
 $\alpha_j(x)$ are the transformation parameters, it is necessary to replace the derivative with a covariant one, adding a triplet of vector fields $\rho^j_{\mu}$,
according to the original idea of Yang and Mills \cite{Yang:1954ek}:
    \begin{equation}
\label{covar-deriv}
\partial_{\mu} \rightarrow D_{\mu} = \partial_{\mu} - \mathbf{i} g_N \rho^j_{\mu}\tau^j.
\end{equation}
Moreover, to ensure the invariance of Lagrangian (\ref{Fermion-lagr}) with covariant derivative
 (\ref{covar-deriv}) under transformation (\ref{gauge-tranf_spinor}), it is necessary that the vector field $\rho^j_{\mu}$ be transformed as follows:
    \begin{equation}
\label{gauge-trans-vector}
\rho_{\mu} \rightarrow \rho^\prime_{\mu} = U\rho_{\mu}U^{-1} -(\partial_{\mu}U)U^{-1},
\end{equation}
where $\rho_{\mu}(x) = \rho^j_{\mu}(x)\tau^j$.
The massless Yang-Mills gauge vector field $\rho^j_{\mu}(x)$ is described by the Lagrangian
    \begin{equation}
\label{Vector-lagr}
L_{\rho} = - \frac{1}{4} \rho^j_{\mu\nu} \rho^{j\,\mu\nu},
\end{equation}
where the vector field strength tensor is given by the well-known formula:
$$ \rho^j_{\mu\nu} = \partial_{\mu}\rho^j_{\nu} - \partial_{\nu}\rho^j_{\mu} + g_N \epsilon^{jkl}
\rho^k_{\mu} \rho^l_{\nu}. $$

By analogy with the BEH mechanism in the Standard Model, we add to our construction a complex scalar field $\Phi_I$, which is a doublet with respect to the gauge group $SU_I(2)$.
We choose the Lagrangian of this field in complete analogy with the Lagrangian of the Higgs field in the SM:
\begin{equation}
L_{\Phi}=D_{\mu}\Phi_I^{\dag} D^{\mu} \Phi_I- {\mu_I}^2\Phi_I^{\dag}\Phi_I -
\lambda_I(\Phi_I^{\dag}\Phi_I)^2,
\label{Scalar-lagr}
\end{equation}
in which only the operators, i.e. the terms of the Lagrangian constructed from the fields, with dimensions not exceeding 4 are kept and the covariant derivative is defined 
by formula (\ref{covar-deriv}).
  In the potential the field  $\varPhi_I$
\begin{equation}
V={\mu_I}^2\Phi_I^{\dag}\Phi_I + \lambda_I(\Phi_I^{\dag}\Phi_I)^2,
\label{Potential}
\end{equation}
the constant $\lambda_I$  must be positive to ensure the stability of the field system, and the potential itself has a non-trivial minimum at negative ${\mu_I}^2 = - |{\mu_I}^2| <0$. Using a gauge transformation, one can pass to the unitary gauge, in which the field $\Phi_I$ takes the following form:
\begin{eqnarray}
\label{scalar field}
\Phi_I(x) = \left(
\begin{array}{c}
0\\
(v_I+ \sigma(x))/\sqrt{2}
\end{array}\right)
\end{eqnarray}
where the vacuum expectation value $v_I$, the condensate, is given by
$$v_I = \sqrt{\frac{|\mu_I^2|}{2\lambda_I}}.$$ For the zero value of the excitation over vacuum, 
the vacuum itself or in other words the vacuum configuration of the field has the form:
\begin{eqnarray}
\label{vacuum}
\Phi^{min}_I(x) = \left(
\begin{array}{c}
0\\
v_I/\sqrt{2}
\end{array}\right).
\end{eqnarray}
The scalar field $\sigma(x)$ in this construction is analogous to the field of the Higgs boson in the SM.

In the effective field theory of nucleon interactions under consideration, the fields $\rho^j(x)$ and $\sigma(x)$ are certain effective fields constructed from fundamental QCD fields in the nonperturbative region.
One can try to identify the vector fields
\begin{equation}
\rho_{\mu}^\pm \equiv \frac{1}{\sqrt{2}}(\rho_{\mu}^1 \mp \mathbf{i} \rho_{\mu}^2), \,\,
\rho_{\mu}^0 \equiv \rho_{\mu}^3
\label{rho-mesons}
\end{equation}
with the fields of $\rho$-mesons. The scalar field $\sigma(x)$ can be the field of the so-called $\sigma$-meson, a field of the glueball type, and the vacuum expectation value $v_I$ can be related
to the characteristic scale of QCD dimensional transmutation $\Lambda_{QCD}$, which in turn is somehow determined by the nonperturbative vacuum gluon condensate. 

As a result of spontaneous symmetry breaking, the Lagrangian of the considered effective theory, consisting of
fermionic (\ref{Fermion-lagr}), vector (\ref{Vector-lagr}) and scalar (\ref{Scalar-lagr}) parts,
takes the following form in terms of the fields $\psi_p$, $\psi_n$, $\rho^\pm$ and $\sigma$:
  \begin{equation}
L = L_{\Psi} + L_{\rho} + L_{\Phi},
\label{Lagr1}
\end{equation}
where
\begin{eqnarray}
\label{LagrPsi}
  L_{\Psi} = \bar\psi_p (\mathbf{i}\partial_{\mu}\gamma^{\mu} - M_N^0)\psi_p
            + \bar\psi_n (i\partial_{\mu}\gamma^{\mu} - M_N^0)\psi_n \\ \nonumber
            + \frac{g_N}{\sqrt{2}}\bar\psi_p\gamma^{\mu}\psi_n\rho_{\mu}^+
            + \frac{g_N}{\sqrt{2}}\bar\psi_n\gamma^{\mu}\psi_p\rho_{\mu}^- \\ \nonumber
            + \frac{g_N}{2}\bar\psi_p\gamma^{\mu}\psi_p\rho_{\mu}^0
            - \frac{g_N}{2}\bar\psi_n\gamma^{\mu}\psi_n\rho_{\mu}^0,
\end{eqnarray}

\begin{eqnarray}
\label{LagrRho}
 L_{\rho} = - \frac{1}{4} \rho^i_{\mu\nu} \rho^{i\,\mu\nu} = \\ \nonumber
- \frac{1}{2}\left(\partial_{\mu}\rho^+_{\nu} - \partial_{\nu}\rho^+_{\mu}\right)
\left(\partial^{\mu}\rho^{-\,\nu} - \partial^{\nu}\rho^{-\,\mu}\right) \\ \nonumber
- \frac{1}{4}\left(\partial_{\mu}\rho^0_{\nu} - \partial_{\nu}\rho^0_{\mu}\right)
\left(\partial^{\mu}\rho^{0\,\nu} - \partial^{\nu}\rho^{0\,\mu}\right) + L_{int},
\end{eqnarray}
the Lagrangian $L_{int}$ containing $\rho$-meson interaction vertices $\rho^+\rho^-\rho^0$
  and $\rho^+\rho^-\rho^0\rho^0$,
completely similar in Lorentzian structure to the interaction vertices of the electroweak $W^\pm$- and $Z$-bosons in the Standard Model, and 
\begin{eqnarray}
\label{LagrSigma}
  L_{\Phi} = \frac{1}{2}\partial_{\mu}\sigma\partial^{\mu}\sigma - \frac{1}{2}M_{\sigma}^2\sigma^ 2
- \lambda_Iv_I\sigma^3 - \frac{\lambda_I}{4}\sigma^4 + \frac{\lambda_I}{4}v_I^4 \\ \nonumber
+ M_{\rho}^2\left(1+\frac{\sigma}{v_I}\right)^2\rho^+_{\mu}\rho^{-\,\mu}
+ \frac{1}{2}M_{\rho}^2\left(1+\frac{\sigma}{v_I}\right)^2\rho^0_{\mu}\rho^{0\, \mu}.
\end{eqnarray}
The masses of the scalar $\sigma$-meson, the vector mesons $\rho^\pm$ and $\rho^0$ in 
(\ref{LagrSigma}) are proportional to the vacuum expectation value $v_I$ and are given by the following relations:
  \begin{equation}
M_{\sigma} = \sqrt{2\lambda_I}v_I
\label{sigma-mass}
\end{equation}
And
  \begin{equation}
M_{\rho^\pm} = M_{\rho^0}\equiv M_{\rho} = \frac{1}{2}g_Nv_I.
\label{rho-mass}
\end{equation}

\section{Gauge-invariant corrections to nucleon masses}

In the Standard Model, there is only one operator of dimension 5 allowed by gauge invariance, the so-called Weinberg operator \cite{Weinberg:1979sa}. In the construction under consideration, the gauge
isotopic invariance admits two operators of dimension 5 due to the vector, and not chiral, as in the SM, nature of the interactions of proton and neutron fields. These two operators give the following addition
to the Lagrangian of the constructed theory:

\begin{eqnarray}
\label{L5}
  L_{5} = \frac{c_1}{\Lambda}\left(\bar\Psi_N \Phi_I \right)\left(\Phi_I^{\dag} \Psi_N\right)
+ \frac{c_2}{\Lambda}\left(\bar\Psi_N \Phi^{c}_I \right)\left(\Phi^{c \,\dag}_I \Psi_N\right),
\end{eqnarray}
where $c_1, c_2$ are some dimensionless constants, $\Lambda$ is a characteristic scale of the order of the nucleon mass, the field
$\Phi^{c} = i \sigma_2\Phi^{\star}$ is the charge conjugate field of the complex scalar doublet $\Phi$.
In the unitary gauge, the Lagrangian (\ref{L5}) takes the form:
\begin{eqnarray}
\label{L5a}
  L_{5} = \frac{v^2_I}{2\Lambda}\left(c_1\bar\psi_n \psi_n + c_2 \bar\psi_p \psi_p\right)
          \left(1+\frac{\sigma}{v_I}\right)^2.
\end{eqnarray}
Note that the gauge invariance allows constants $c_1$ and $c_2$ that are not equal to each other, which leads to the fact that the additions to the masses of the lower and upper fermions, i.e. to the masses of neutrons and protons will be different:
\begin{eqnarray}
\label{neutron-proton-masses}
m_n = M_N + c_1\frac{v^2_I}{2\Lambda} \\ \nonumber
m_p = M_N + c_2\frac{v^2_I}{2\Lambda}.
\end{eqnarray}
We emphasize that, in this way, the difference in the proton and neutron masses can arise without violating the basic principle of gauge invariance with respect to the $SU_I(2)$ group and not be related to the difference in
electric charges of the proton and neutron. Note also that the characteristic large parameter
in this effective field theory  is the nucleon mass, i.e. the scale of $\Lambda$ must be of the order of $M_N$.

The Lagrangian (\ref{L5}) implies the Lagrangian for the interaction of the boson scalar field $\sigma$ with the proton and neutron fermionic fields, which, in addition to the Yukawa-type interaction, also contains a term proportional to the square of the field $\sigma$ :
  \begin{eqnarray}
\label{Yukawa}
  L_{\bar{\psi} \psi \sigma} =
\frac{v_I}{\Lambda} \left(c_1\bar\psi_n \psi_n + c_2 \bar\psi_p \psi_p\right) \sigma \\ \nonumber
+ \frac{1}{2\Lambda} \left(c_1\bar\psi_n \psi_n + c_2 \bar\psi_p \psi_p\right) \sigma^2.
\end{eqnarray}

\section{Electromagnetic interactions}

As is well known, the electromagnetic interaction violates isotopic invariance. Electromagnetic interactions can be introduced into the considered effective theory again using the analogy with the construction of the Standard Model. We assume that the hypercharge group $U_Y(1)$ is also a gauge group and require that the Lagrangian be invariant under the gauge group $SU_I(2)\times U_Y(1)$.
We choose the Lagrangian of such a theory in the form similar to (\ref{Lagr1})
\begin{equation}
L = L_{\Psi} + L_{\rho} + L_{B^Y}+ L_{\Phi} ,
\label{Lagr2}
\end{equation}
where in the fermionic Lagrangian $L_{\Psi}$ (\ref{Fermion-lagr}) and in the scalar Lagrangian
$L_{\Phi}$ (\ref{Scalar-lagr}) the covariant derivative contains an additional term related to the vector gauge field $B^Y_{\mu}$ of the hypercharge group $U_Y(1)$ :
\begin{equation}
\label{covar-deriv-2}
D_{\mu} = \partial_{\mu} - \mathbf{i} g_N \rho^j_{\mu}\tau^j - \mathbf{i} g_Y \frac{Y}{2} B^Y_{\mu},
\end{equation}
where $Y$ is the hypercharge of either the nucleon field $\Psi_N$ or the field of the complex scalar doublet $\Phi_I$.
The Lagrangian $L_{\rho}$ (\ref{Vector-lagr}) of the non-Abelian vector field $\rho(x)$
is supplemented by 
the Lagrangian of the Abelian vector field $B^Y_{\mu}$:
   \begin{equation}
\label{Vector-lagr-B}
L_{B^Y} = - \frac{1}{4} B^Y_{\mu\nu} B^{Y\,\mu\nu},
\end{equation}
where the field strength tensor $B^Y_{\mu}$ has the well-known form
$$ B^Y_{\mu\nu} = \partial_{\mu}B^Y_{\nu} - \partial_{\nu}B^Y_{\mu}.$$

In case of spontaneous symmetry breaking, similarly to the Standard Model, the symmetry of the $SU_I(2)\times U_Y(1)$ theory under consideration is broken up to the symmetry of the electromagnetic group 
$U(1)_{em}$. We note that this violation leaves the global invariance $U(1)$, which corresponds to the conservation of the baryon charge.

Acting by analogy with the SM, we introduce the same charged
gauge fields as in (\ref{rho-mesons})
\begin{equation}
\rho_{\mu}^\pm \equiv \frac{1}{\sqrt{2}}(\rho_{\mu}^1 \mp \mathbf{i} \rho_{\mu}^2).
\label{rho-pm}
\end{equation}
Charged fields (\ref{rho-pm}) are described by the Lagrangians constructed above (see (\ref{LagrPsi})
  and (\ref{LagrRho})). In particular, the Lagrangian of charged currents of protons and neutrons is determined by the second line of formula (\ref{LagrPsi}) and has the following form:
\begin{equation}
L_{CC} = \frac{g_N}{\sqrt{2}}\bar\psi_p\gamma^{\mu}\psi_n\rho_{\mu}^+
            + \frac{g_N}{\sqrt{2}}\bar\psi_n\gamma^{\mu}\psi_p\rho_{\mu}^-.
\label{LCC}
\end{equation}

Unlike the neutral field in (\ref{rho-mesons}), again by analogy with the SM, the neutral fields $\rho_{\mu}^3$ and $B^Y_{\mu}$ can be expressed in terms of an orthogonal combination of two new  physical vector fields $\rho_{\mu}^0$ and $A_{\mu}$, which will have certain masses :
\begin{eqnarray}
\label{rho-A}
\rho_{\mu}^3 \equiv \rho_{\mu}^0 \cos{\theta_I} + A_{\mu}\sin{\theta_I} \\ \nonumber
B^Y_{\mu} \equiv -\rho_{\mu}^0 \sin{\theta_I} + A_{\mu}\cos{\theta_I}.
\end{eqnarray} 

The requirement for the correct electromagnetic interaction of the $\psi_p$ and $\psi_n$ fermionic fields with the $A_{\mu}$ field, which is associated with the photon field, leads to an equality following from Lagrangian (\ref{Fermion-lagr}), taking into account the replacement of the usual derivative with the covariant one, defined by the expression (\ref{covar-deriv-2}):
\begin{eqnarray}
\label{eqn-A-psi}
g_N T^3_{\psi_{p,n}} \sin{\theta_I} + g_Y \frac{Y}{2}\cos{\theta_I} = e Q_{\psi_{p,n}},
\end{eqnarray}
where the electric charges and projections of the isotopic spins of the proton and neutron are $Q_{\psi_{p}} = 1$ and $Q_{\psi_{n}} = 0$, $T^3_{\psi_{p}} = \frac{1}{2}$ and
$T^3_{\psi_{n}} = -\frac{1}{2}$, and the constant $e$ is related to the fine structure constant $\alpha$ by the well-known equality $\alpha = \frac{e^2} {4\pi}$. The Gell-Mann-Nishijima relation $Q = \frac{Y}{2} + T^3$ for the proton $\psi_{p}$, neutron $\psi_{n}$ and $\Phi_I$ fields gives the hypercharge value $Y = 1$. Taking into account this ratio, from  equation (\ref{eqn-A-psi}) we get:
\begin{eqnarray}
\label{eqn-for-coupling}
g_N \sin{\theta_I} = e \\ \nonumber
g_Y \cos{\theta_I} = e
\end{eqnarray}
Using equalities (\ref{eqn-for-coupling}), the Lagrangian for the interaction of protons and neutrons with the neutral field $\rho_{\mu}^0$ takes the following form:
\begin{eqnarray}
\label{Lagr-NC-rho}
 L_{NC} = 
\bar\psi_p \rho_{\mu}^0 \gamma^{\mu}\psi_p \frac{g_N}{2}
\frac{1-2e^2/g_N^2}{\sqrt{1-e^2/g_N^2}}  
\,+\,\bar\psi_n \rho_{\mu}^0\gamma^{\mu}\psi_n \frac{-g_N}{2}
\frac{1}{\sqrt{1-e^2/g_N^2}} 
\end{eqnarray}
  It can be seen that for $e^2 \ll g_N^2$ Lagrangian (\ref{Lagr-NC-rho}) passes into the corresponding part of the Lagrangian (\ref{LagrPsi}).

Note that the masses of the mesons  $\rho^\pm$ and $\rho^0$, which follow from the Lagrangian $L_{\Phi}$ after spontaneous symmetry breaking, are expressed by a formula similar to the relation between the masses of the  $W^\pm $- and $Z$-bosons:
  \begin{eqnarray}
\label{eqn-for-mass}
M_{\rho^\pm} = M_{\rho^0}\cos{\theta_I}= M_{\rho^0}\sqrt{1-e^2/g_N^2}.
\end{eqnarray}
Thus, this theory predicts the difference between the masses of $\rho^\pm$- and $\rho^0$-mesons
  \begin{eqnarray}
\label{eqn-for-mass-diff}
M_{\rho^0} - M_{\rho^\pm} \simeq M_{\rho^0}\frac{e^2}{2g_N^2}=M_{\rho^0}\frac{\alpha }{2\alpha_N},
\end{eqnarray}
where $\alpha_N = \frac{g_N^2}{4\pi}$.
If we require that this difference does not exceed a value of the order of $1 MeV$
\cite{ParticleDataGroup:2022pth}, then, taking into account the known values of the fine structure constant and the mass of the $\rho^0$ meson, we obtain a lower bound greater than one for the constant
 $\alpha_N$:
  \begin{eqnarray}
\label{alpha-N}
\alpha_N \geqslant 1.
\end{eqnarray}
This means that the constructed theory is in the strong coupling regime.  

\section{Gauge-invariant effective interactions involving pions}

As mentioned above, $\pi$-mesons (pions) appear as pseudo-Goldstone bosons when the $SU(2)_L\times SU(2)_R \rightarrow SU(2)_I$ chiral symmetry is broken. The gauge invariance $SU(2)_I\times U(1)_Y$ determines the structure of interactions of pions as isotopic spin triplets with the proton, neutron, $\rho^\pm$-, $\rho^0$ and $\sigma$ mesons and photons. The pion interaction Lagrangian for operators of dimension $\leq 4$, dictated by the principle of gauge invariance, has the form:
\begin{eqnarray}
\label{Lagr-pion}
 L_{\pi} = 
\frac{1}{2} \left(D_{\mu}^{ij} \pi^j\right) \left(D^{\mu \,ik} \pi^k\right)
 - \frac{1}{2} m_{\pi}^{0\, 2} \pi^i \pi^i \\ \nonumber
+ \lambda_{\pi\sigma} \pi^i \pi^i \Phi_I^{\dag}\Phi_I 
+ i g_{\pi NN} \bar\Psi_N \tau^i \gamma^5 \Psi_N \pi^i
\\ \nonumber
+ (higher \,\, dimensional \,\,terms),
\end{eqnarray}
where the indices $i,j,k = 1,2,3$. Taking into account that the pion field hypercharge $Y_{\pi}$ is equal to zero, due to the relation $Q_{\pi} = \frac{Y_{\pi}}{2} + T_{3\, \pi}$, the covariant derivative is defined as:
\begin{eqnarray}
\label{covar-deriv-pion}
D_{\mu}^{ij} = \partial_{\mu} \delta^{ij} - \mathbf{i} g_N \epsilon^{ijk} \rho_{\mu}^k.
\end{eqnarray}
In the unitary gauge using obvious field redefinitions
\begin{eqnarray}
\label{rho-pi}
\rho_{\mu}^\pm \equiv \frac{1}{\sqrt{2}}(\rho_{\mu}^1 \mp \mathbf{i} \rho_{\mu}^2), \,\,
\rho_{\mu}^3 \equiv \rho_{\mu}^0 \cos{\theta_I} + A_{\mu}\sin{\theta_I} \\ \nonumber
\pi^\pm \equiv \frac{1}{\sqrt{2}}(\pi^1 \mp \mathbf{i} \pi^2), \,\, \pi^3 \equiv \pi^0
\end{eqnarray}
the Lagrangian of pion interactions (\ref{Lagr-pion}) takes the following form:
\begin{eqnarray}
\label{Lagr-pi-component}
L_{\pi} = 
\partial_{\mu}\pi^+\partial^{\mu}\pi^- - m_{\pi}^2 \pi^+ \pi^-
+ \frac{1}{2}\partial_{\mu}\pi^0\partial^{\mu}\pi^0  - \frac{1}{2} m_{\pi}^2 \pi^0 \pi^0 \\\nonumber 
+ \lambda_{\pi\sigma} \left( 2 v_I \pi^+ \pi^- \sigma + \pi^+ \pi^- \sigma^2 + v_I \pi^0 \pi^0 \sigma
+\frac{1}{2} \pi^0 \pi^0 \sigma^2 \right)\\ \nonumber
+ i \frac{1}{2}g_{\pi NN}\left( \bar\psi_p \gamma^5 \psi_n \pi^+ + \bar\psi_n \gamma^5 \psi_p \pi^- 
+ \bar\psi_p \gamma^5 \psi_p \pi^0 - \bar\psi_n \gamma^5 \psi_n \pi^0\right)\\ \nonumber
+ g_N \left(  \partial_{\mu}\pi^- \pi^+ - \partial_{\mu}\pi^+ \pi^-\right)
      \left( \rho^{0 \, \mu}\cos\theta_I + A^{\mu}\sin \theta_I \right) \\ \nonumber
+ g_N \left[\left( \partial_{\mu}\pi^- \pi^0 - \pi^-\partial_{\mu}\pi^0\right)\rho^{+ \, \mu}
           -\left( \partial_{\mu}\pi^+ \pi^0 - \pi^+\partial_{\mu}\pi^0\right)\rho^{-\, \mu} \right]
 \\ \nonumber
+ g_N^2 \left( 2 \pi^+\pi^-\rho^{+ \, \mu}\rho^-_{\mu} - \pi^-\pi^-\rho^{+ \, \mu}\rho^+_{\mu} 
             - \pi^+\pi^+\rho^{-\, \mu}\rho^-_{\mu}
               + 2 \pi^0\pi^0\rho^{+ \, \mu}\rho^-_{\mu}\right)\\ \nonumber
+ g_N^2 \left(2\pi^+\pi^- + \pi^0\pi^0 \right)
        \left( \rho^{0}_{\mu}\cos\theta_I + A_{\mu}\sin \theta_I \right)
         \left( \rho^{0 \, \mu}\cos\theta_I + A^{\mu}\sin \theta_I \right), \nonumber
\end{eqnarray} 
where $m_{\pi}^2 = m_{\pi}^{0\, 2} - \lambda_{\pi\sigma} v_I^2$, i.e. the bare mass of the pion receives a correction due to the interaction with the scalar field. The sign of the constant $\lambda_{\pi\sigma}$ can be either positive or negative. The Lagrangian (\ref{Lagr-pi-component}) determines the structure of the interaction vertices of pion fields $\pi^\pm, \pi^0$ with fields $\rho^\pm, \rho^0, A, \sigma, \ psi_{p}, \psi_{n}$. 

The Lagrangians (\ref{Lagr2}) and (\ref{Lagr-pi-component}) contain various triple and quartic field interaction vertices shown in the figures below. Figure 1 shows interaction vertices involving $\rho$-mesons. Figure 2 shows interaction vertices involving $\sigma$ and $\rho$ mesons. The vertices of interaction with the participation of pi-mesons are shown in Fig.3.\\

\begin{center}
\includegraphics{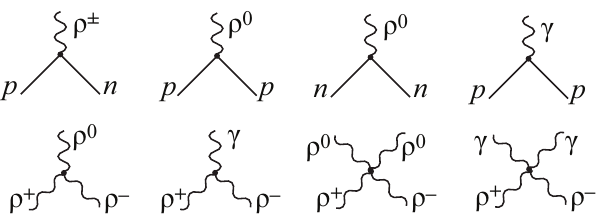}\\
{\footnotesize Fig.1 Interaction vertices involving $\rho$-mesons.}
\\
\vspace{40pt}
\includegraphics{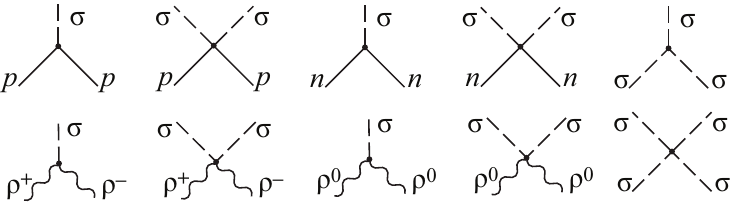}\\
{\footnotesize Fig.2 Interaction vertices involving $\sigma$- and $\rho$-mesons.}\\
\vspace{40pt}
\includegraphics{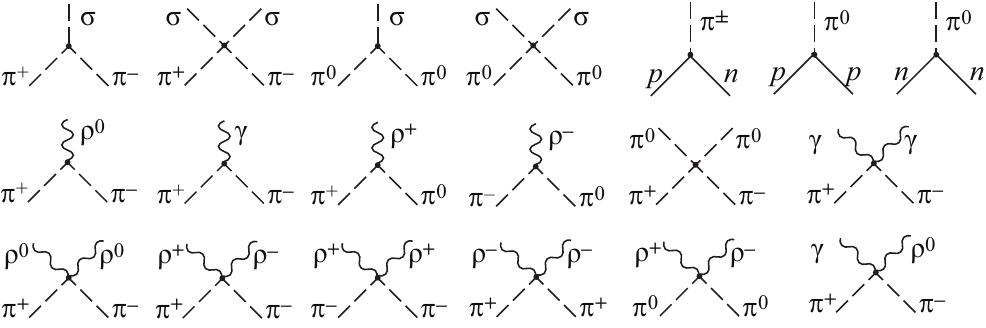}\\
{\footnotesize Fig.3 Interaction vertices involving pions.}\\
\end{center}


\section{Meson Decays and Effective Theory Parameters}

The Lagrangian (\ref{Lagr-pi-component}) contains, among other things, the triple vertices of the interaction of $\rho$-mesons with pi-mesons, shown in the second line of Fig.3. The Feynman diagrams corresponding to these vertices describe the decay processes
  $\rho^{0} \rightarrow \pi^+\pi^-, \, \rho^{\pm} \rightarrow \pi^{\pm} \pi^0$.
The width of each of these decays is given by the same formula:
\begin{eqnarray}
\label{Rho-width}
\Gamma_\rho = \frac{g_N^2}{48\pi}M_\rho \beta^3,
\end{eqnarray}
where $\beta = \sqrt{1-4m_{\pi}^2/M_{\rho}^2}$, $M_{\rho}$ is the mass of the decaying $\rho$ meson,
either $\rho^{0}$ or $\rho^{\pm}$. It follows from the formula (\ref{Rho-width}) that in order for the width of the $\rho$ meson to be approximately equal to $\Gamma_\rho \simeq 147 MeV$ for the masses of the $\rho$ meson $M_\rho \simeq 770$\,MeV and pi-meson $m_{\pi} \simeq 139.6$\,MeV the coupling constant should be about
$g_N \simeq 5.97$ and, accordingly,
\begin{eqnarray}
\label{alfa-N-numerical}
\alpha_N = \frac{g_N^2}{4\pi} \simeq 2.84,
\end{eqnarray}
which is quite consistent with the estimate (\ref{alpha-N}). The resulting value of constant $\alpha_N$
(\ref{alfa-N-numerical}), due to the formula (\ref{eqn-for-mass-diff}), leads to the following value of the mass difference between the neutral and charged $\rho$ mesons:
\begin{eqnarray}
\label{eqn-for-mass-diff-num}
M_{\rho^0} - M_{\rho^\pm} \simeq 0.989 \, MeV,
\end{eqnarray}
From formula (\ref{rho-mass}) relating the mass of the $\rho^0$ meson to  the vacuum expectation value $v_I$,
the latter can be found to be:
\begin{equation}
v_I = 2 M_{\rho^0}/g_N \simeq 260 \, MeV.
\label{vacuum-expectation}
\end{equation}
We note that numerically vacuum expectation value (\ref{vacuum-expectation}) corresponds to the characteristic QCD scale $\Lambda_{QCD}$. As can be seen, for such a  vacuum expectation value, the expansion of the effective theory in terms of the ratio of the vacuum expectation parameter to the characteristic scale of the effective theory $M_N\simeq 1$\,GeV 
  \begin{equation}
v_I/M_N\simeq 0.26.
\label{expansion-parameter}
\end{equation}
makes sense.

The mass of the $\sigma$ meson, which is also called the $f_0(500)$ meson \cite{ParticleDataGroup:2022pth}, is determined with a rather poor accuracy. If we choose a value for this mass in the interval
  \begin{equation}
M_{\sigma} = 400 - 500 \, MeV,
\label{sigma-mass-num}
\end{equation}
then from formula (\ref{sigma-mass}) relating the mass of the $\sigma$-meson to the vacuum expectation value, the value for the coupling constant $\lambda_I$ is obtained in the interval:
  \begin{equation}
\lambda_I = 1.2 - 1.8 .
\label{sigma-coupling}
\end{equation}
From the form of the interaction vertex of the $\sigma$ field with the fields of the charged and neutral pions following from the Lagrangian (\ref{Lagr-pi-component}), it is easy to obtain an expression for the width of the $\sigma$-meson decay into a pair of pions:
\begin{eqnarray}
\label{sigma-width}
\Gamma_\sigma = \frac{3}{8\pi} \lambda_{\pi\sigma}^2 \frac{v_I^2}{M_{\sigma}} \beta,
\end{eqnarray}
where $\beta = \sqrt{1-4m_{\pi}^2/M_{\sigma}^2}$. In this case, the decay branchings in the charged and neutral modes are:
  \begin{eqnarray}
\label{sigma-branching}
Br(\sigma \rightarrow \pi^+\pi^-) = 2/3, \,\,\, Br(\sigma \rightarrow \pi^0\pi^0) = 1/3.
\end{eqnarray}
The $\sigma$-meson can decay into two photons through triangular diagrams involving a proton and a charged $\rho$-meson , similarly to the decay of the Higgs boson into two photons in the SM. But the decay  branching ratio into two photons is small, and we do not present it here.

If, for definiteness, we choose the mass of the $\sigma$-meson 500\,MeV, and the decay width of the $\sigma$-meson, which is also determined with a poor accuracy (see \cite{ParticleDataGroup:2022pth}), choose about 400\,MeV , then the formula for the decay width (\ref{sigma-width}) implies an estimate for the coupling constant $\lambda_{\pi\sigma}$ in the Lagrangian (\ref{Lagr-pion}):
   \begin{equation}
\lambda_{\pi\sigma} \simeq 5.5 .
\label{sigma-pi-coupling}
\end{equation}
Such a value of the constant also indicates the strong coupling mode of the considered effective theory.

Taking into account the found  vacuum expectation value (\ref{expansion-parameter}) and the well-known masses of protons and neutrons, one can obtain approximate values for the coefficients $c_1$ and $c_2$ in the mass formula (\ref{neutron-proton-masses}) , if, for definiteness, we choose the values of the parameter $M_N = \Lambda = 1$\,GeV:
\begin{eqnarray}
\label{neutron-proton-masses-diff}
c_1 = (m_n - M_N)\frac{2\Lambda}{v^2_I} \simeq -1.79 \\ \nonumber
c_2 = (m_p - M_N)\frac{2\Lambda}{v^2_I} \simeq -1.83.
\end{eqnarray}
The coefficients $c_1$ and $c_2$ for operators of dimension 5 (see (\ref{L5})) turn out to be of the order of unity, i.e. natural order of magnitude when expanded in terms of the small parameter $v_I/M_N$ (\ref{expansion-parameter}).

The interaction vertices that appear in the constructed theory, shown in Figs.1,2,3, allow constructing Feynman diagrams that contribute to the potential of nucleon-nucleon interaction with the exchange of one boson (one-boson exchange potential), as shown in the following figures (Figs.4,5) .

\begin{center}

\includegraphics{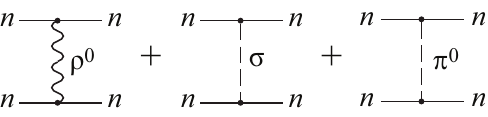}\\
{\footnotesize Fig.4 One-boson exchange diagrams for $n-n$-interaction.}\\
\vspace{20pt}
\includegraphics{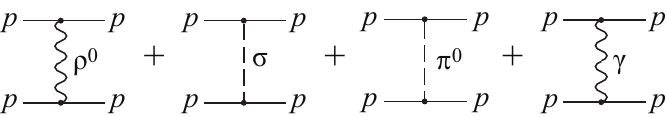}\\
{\footnotesize Fig.5 One-boson exchange diagrams for $p-p$-interaction.}\\
\vspace{20pt}

\includegraphics{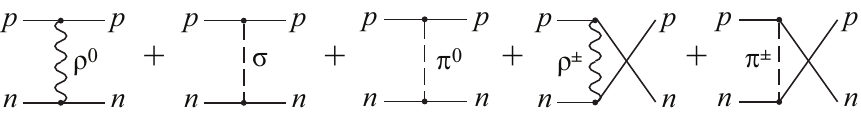}\\
{\footnotesize Fig.6 One-boson exchange diagrams for $p-n$-interaction.}\\
\end{center} 

Note that the triple and quadruple vertices lead to irreducible contributions to the three-particle and four-particle nucleon interactions. Moreover, due to the fact that the gauge constant
$g_N$ between nucleons and $\rho$-mesons enters the four-particle vertices quadratically, and, being noticeably greater than unity, significantly enhances the four-particle nucleon interactions.
This remark requires a separate detailed consideration. Note that the Feynman rules for the triple and quadruple vertices of the gauge bosons $\rho^0$, $\rho^\pm$, $\gamma$, up to an obvious renaming of the coupling constants, correspond to the vertices with $Z$-, $W^\pm $-bosons and $\gamma$ in the SM. 

The diagrams in Figs.4,5,6 also describe the amplitudes of the processes of nucleon-nucleon scattering, while taking into account all the relative signs of the diagrams in the amplitude, and, accordingly, all the interference contributions to the squares of the matrix elements and the cross section of the processes. In calculations, however, it must be kept in mind that the range of allowable energies at which the effective model operates must be limited from above by the characteristic scale $\Lambda$:
\begin{eqnarray}
\label{Energy-intreval}
E \,\, < \,\, \Lambda \,\simeq \,M_N
\end{eqnarray}

\section{Concluding remarks}

The effective field theory based on the idea of treating isotopic and hypercharge symmetries as a local gauge invariance with gauge group $SU(2)_I\times U(1)_Y$ is constructed. This symmetry is spontaneously broken down to the electromagnetic group $U(1)_{em}$ with the help of a complex scalar doublet with a nontrivial vacuum expectation value, an analog of the Higgs field in the Standard Model. The found vacuum expectation value turns out to be quite close to the value of the dimensional transmutation parameter, $\Lambda_{QCD}$, reflecting the complex structure of the QCD vacuum. Of course, the question of the relationship between the Lagrangian of the constructed effective field theory and the fundamental theory of strong interactions, quantum chromodynamics, remains open.

The Lagrangian of the theory consists of parts represented in the Lagrangians (\ref{Lagr2}) and (\ref{Lagr-pion}), which include operators of dimension not higher than 4, and also represented in the Lagrangian (\ref{L5}), which includes the dimension 5 operators. It is shown how these Lagrangians are expressed in terms of the physical fields. Note that adding gauge-invariant operators of higher dimensions to the Lagrangian, in our effective theory the operators of dimension 5, along with operators of dimension $\leq 4$, which were constructed by analogy with the SM, is carried out by analogy with the Standard Model Effective Field Theory approach (SMEFT) \cite{Weinberg:1979sa, Buchmuller:1985jz, Grzadkowski:2010es} (see also reviews \cite{Boos:2022cys, Dawson:2022ewj}).

The vertices of the interaction of protons, neutrons, $\rho$-, $\sigma$-mesons, pions and photons following from these Lagrangians are graphically shown in Figs.1,2,3. The main gauge coupling constant $\alpha_N$ is estimated from the decay width of $\rho$ mesons into pions and is approximately $2.84$ (\ref{alfa-N-numerical}), which is noticeably larger than 1.  It follows from the constructed theory that the masses of the neutral and charged $\rho$ mesons should differ slightly, and this difference (\ref{eqn-for-mass-diff-num}) is about $1\,MeV$. The vacuum expectation value, determined by the $\rho$-meson mass and the gauge coupling constant, is approximately $v_I \simeq 260 \, MeV$ (\ref{vacuum-expectation}). Thus, the expansion parameter of the effective field theory $v_I/M_N$ is rather small, which makes it possible to construct a perturbation theory on this parameter. In particular, two gauge operators of dimension 5 make a parametrically small contribution to the neutron mass and to the proton mass, as follows from (\ref{L5a}). Thus, it becomes possible, by choosing different coefficients in front of the operators, to obtain the difference in the proton and neutron masses (\ref{neutron-proton-masses-diff}) without violating the gauge symmetry.

The diagrams shown in Figs.4,5 contribute to the one-boson-exchange potential of the nucleon-nucleon interaction predicted in the theory. Qualitatively, the potential corresponds to the widely discussed internucleon one-boson-exchange potentials (OBEP) (see CD-Bonn potential
 \cite{Machleidt:2000ge} and review \cite{Machleidt:2022kqp}). The constructed field theory also predicts 3- and 4-particle interactions based on 3- and 4-point irreducible interaction vertices of $\rho$-, $\sigma$-mesons, pons, as seen from the Lagrangians (\ref{LagrSigma}) , (\ref{Yukawa}), (\ref{Lagr-NC-rho}), (\ref{Lagr-pi-component}) and shown in the last lines of Figs.1,2,3. 

The vector $\omega$-meson ($\omega_{\mu}$) has no isotopic spin and neither hypercharge nor electric charge. Thus, the $\omega$-meson is sterile with respect to the considered gauge group. An interaction
of the $\omega$-meson with protons and neutrons in the framework of discussed effective theory can be introduced by means of higher dimensional operator as demonstrated in Appendix A.  

The potentials can be obtained from the amplitudes given in Figs.4,5,6 by a well-known method, which is presented in detail in the appendices to review\cite{Machleidt:2022kqp} for various variants of bosons exchanged between nucleons. The explicit form of the potentials following from the exchange diagrams of the presented theory is not given in this paper.
The derivation of symbolic expressions for the amplitudes and nucleon-nucleon potentials, the calculation of scattering cross sections and phase shifts, as well as the corresponding phenomenological analysis and comparison of the results with experimental data go beyond the scope of the presented theoretical study and are planned as next steps.

\section*{Acknowledgments}
The author is grateful to I.P.~Volobuev, D.A.~Lanskoi, T.Yu.~Tretyakova, and A.M.~Shirokov for discussions and valuable remarks.
The work was supported by grant 22-12-00152 of the Russian Science Foundation.


\appendix
\section{Nucleon interaction with the omega-meson}

Since the vector $\omega$-meson ($\omega_{\mu}$) has no isotopic spin, neither hypercharge,
 nor electric charge, it is sterile with respect to the gauge group.
In the absence of interaction, it is described by the well-known Lagrangian of a free massive vector field:
\begin{eqnarray}
\label{omega-lagr}
L_{\omega} = -\frac{1}{4} \omega_{\mu\nu}\omega^{\mu\nu} + \frac{1}{2} M^2_{\omega}\, \, \omega_{\mu} \omega^{\mu},
\end{eqnarray}
where $\omega_{\mu\nu} = \partial_{\mu}\omega^{\nu} -\partial_{\nu}\omega^{\mu}$ is the 
field strength tensor of $\omega_{\mu}$, $M_{\omega}$ is its mass.

A Lagrangian of interaction with the nucleon field can be introduced using a gauge-invariant operator of dimension 5:
\begin{eqnarray}
\label{omega-nuclon}
L_{\bar{\Psi} \Psi \omega } = \frac{g_\omega}{\Lambda}\bar\Psi_N \sigma_ {\mu\nu}\Psi_N\omega^{\mu\nu},
\end{eqnarray}
where $\sigma_ {\mu\nu} = \frac{1}{2}\left(\gamma_\mu \gamma_\nu - \gamma_\nu \gamma_\mu \right)$,
$\gamma_\mu$ being the Dirac matrices. Substituting  the expression for the nucleon
doublet
$\Psi_N(x) = \left(
\begin{array}{c}
\psi_p \\
\psi_n
\end{array}
\right)$
into Lagrangian (\ref{omega-nuclon}), we obtain the Lagrangian for the interaction of the $\omega$-meson with protons and neutrons:
\begin{eqnarray}
\label{omega-proton-neutron}
L_{\bar{\psi} \psi \omega } = \frac{g_\omega}{\Lambda}\left(\bar\psi_p \sigma_ {\mu\nu}\psi_p +
\bar\psi_n \sigma_ {\mu\nu}\psi_n\right)\omega^{\mu\nu}.
\end{eqnarray}
We emphasize that the gauge invariance of Lagrangian (\ref{omega-nuclon}) leads to the fact that protons and neutrons interact with the $\omega$-meson in the same way, the interaction vertices have a tensor structure and are proportional to the same coupling constant $g_\omega$.
Feynman rules following from Lagrangian (\ref{omega-proton-neutron}) give the structure of internucleon 
$\omega$-meson exchange amplitude. The corresponding amplitude could be added to the amplitudes given in Figs.4,5.

\bibliography{isospin-corrected.bib}
\bibliographystyle{bib.bst}

\end{document}